\documentclass{vgtc}                          

\graphicspath{{figures/}{pictures/}{images/}{./}} 

\usepackage{times}                     

\usepackage{tabu}                      
\usepackage{booktabs}                  
\usepackage{lipsum}                    
\usepackage{mwe}                       

\usepackage{mathptmx}                  
\usepackage{multirow}

\onlineid{1171}

\vgtccategory{Poster}

\vgtcinsertpkg

\title{A Collaborative Extended Reality Prototype for \\3D Surgical Planning and Visualization}

\author{Shi Qiu, Ruiyang Li, Qixuan Liu, Yuqi Tong, Yue Qiu, Yinqiao Wang, Yan Li, Chi-Wing Fu, Pheng-Ann Heng\\ %
        \scriptsize Department of Computer Science and Engineering, The Chinese University of Hong Kong\\
        \scriptsize Institute of Medical Intelligence and XR, The Chinese University of Hong Kong\\
        \scriptsize \{shiqiu, liry, qxliu, yqtong, yqiu23, yqwang, yanli, cwfu, pheng\}@cse.cuhk.edu.hk}

\teaser{
  \centering
  \includegraphics[width=0.85\linewidth]{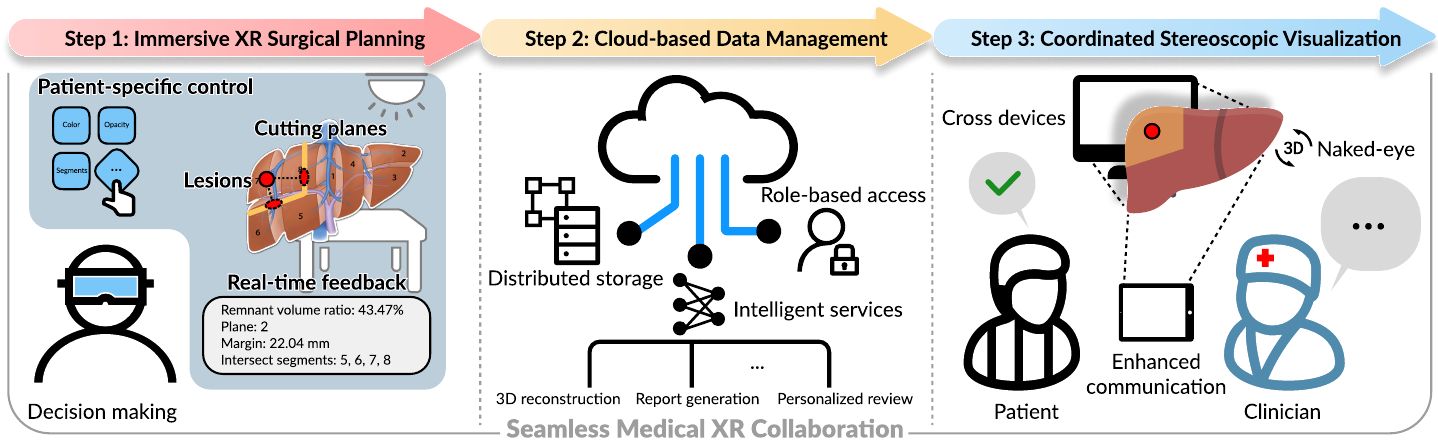}
  \caption{Overview of our system workflow, illustrated using a liver resection surgical planning example.}
  \label{fig:teaser}
}

\abstract{
    We present a collaborative extended reality (XR) prototype for 3D surgical planning and visualization. Our system consists of three key modules: XR-based immersive surgical planning, cloud-based data management, and coordinated stereoscopic 3D displays for interactive visualization. We describe the overall workflow, core functionalities, implementations and setups. By conducting user studies on a liver resection surgical planning case, we demonstrate the effectiveness of our prototype and provide practical insights to inspire future advances in medical XR collaboration. 
} 


\begin{document}


\firstsection{Introduction}

\maketitle


Advances in medical imaging have transformed medical procedures, but everyday clinical practice still depends primarily on 2D displays for viewing CT and MRI scans. Interpreting complex anatomy from stacked slices requires substantial expertise and mental reconstruction, which can lead to overlooked subtle spatial relationships that are critical for surgical planning. Extended reality (XR) offers a promising alternative by enabling immersive, interactive exploration of 3D medical data and supporting more intuitive spatial understanding in visualization~\cite{qiu2025cvhslicer}. However, integrating XR into real clinical workflows remains challenging due to fragmented data pipelines, limited interoperability, lack of real-time coordination, and practical constraints on accessible visualization~\cite{liu2025coor}. Most existing solutions address only isolated aspects of the workflow and lack effective integration for seamless XR collaboration. 

We present a collaborative XR prototype for immersive 3D surgical planning and visualization. Our system comprises three tightly integrated modules: (1) an XR-based immersive planning environment that enables clinicians to intuitively explore patient-specific 3D models; (2) a cloud-based data management platform for secure storage, synchronization, and sharing of surgical plans and associated data; and (3) coordinated stereoscopic 3D displays that deliver interactive visualization on multiple devices without specialized eyewear. Together, our system supports effective medical collaboration, informs clinical decision-making, and enhances communication between clinicians and patients.

\section{System Details}
The overall workflow of our system (Fig.~\ref{fig:teaser}) begins with XR-based surgical planning, where clinicians intuitively explore patient data in an immersive 3D environment. Then, determined surgical plans and relevant data are securely stored and managed through a cloud platform, enabling seamless access and sharing. Finally, coordinated stereoscopic visualization allows users to view and interact with data on multiple 3D displays, facilitating collaborative review and discussion across different devices and locations.
\begin{figure}[h]
    \centering
\includegraphics[width=0.85\columnwidth]{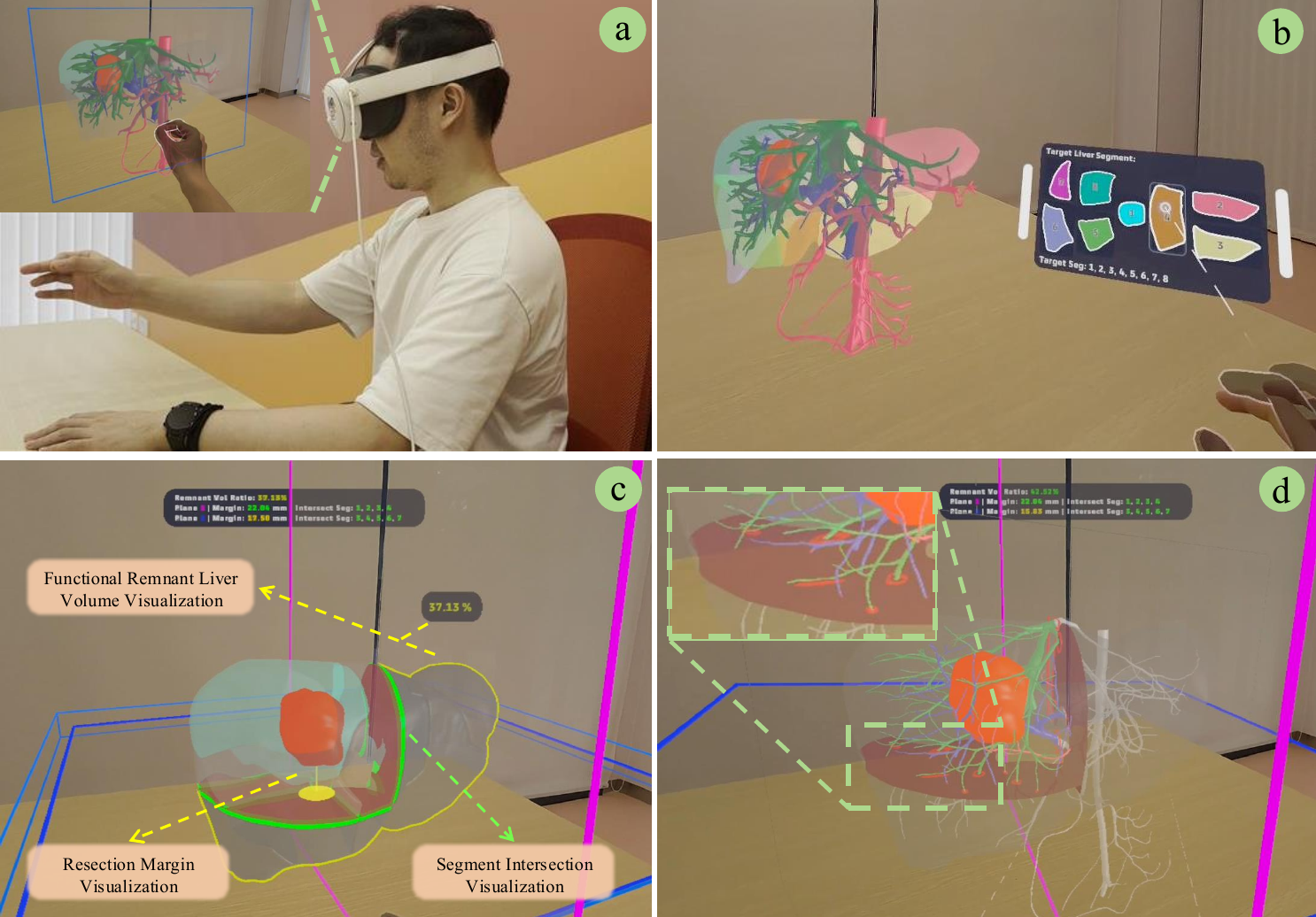} 
    \caption{XR-based surgical planning of liver resection. (a) Overview of the setup with an MR headset. (b) Mesh-based visualization of patient anatomy and the user interface panel. (c) Real-time computation of surgical safety metrics during plan determination. (d) Visualization of intersections between vessels and resection planes.}  \label{fig:surgical_planning_single_col}
\end{figure}

\textbf{Surgical Planning.} We develop a virtual surgical planning environment that runs on MR headsets. The system supports mesh-based anatomical visualization of 3D medical data, such as CT and MRI volumes, providing clinicians with an intuitive and detailed view of patient-specific anatomy. We implement freehand interactions for real-time planning and simulation, enabling users to manipulate resection planes and visualize predicted resection metrics, as illustrated in Fig.~\ref{fig:surgical_planning_single_col}. Leveraging the high-quality see-through and casting capabilities of MR devices, multiple users and observers can collaborate and communicate within the same environment, facilitating efficient teamwork and timely decision-making.

\textbf{Data Management.} We employ a cloud-based platform for the secure and efficient management of patient data and associated surgical plans. Built on the ThinkPHP application framework coupled with the user role-based access control, the platform comprises key data storage, processing, and sharing components, ensuring that sensitive medical data are securely maintained, efficiently optimized for usability, and seamlessly accessed by authorized users across different devices. This data management approach supports real-time collaborative exploration of surgical plans while maintaining robust data security and integrity.
\begin{figure}[h]
    \centering
    \includegraphics[width=.85\columnwidth]{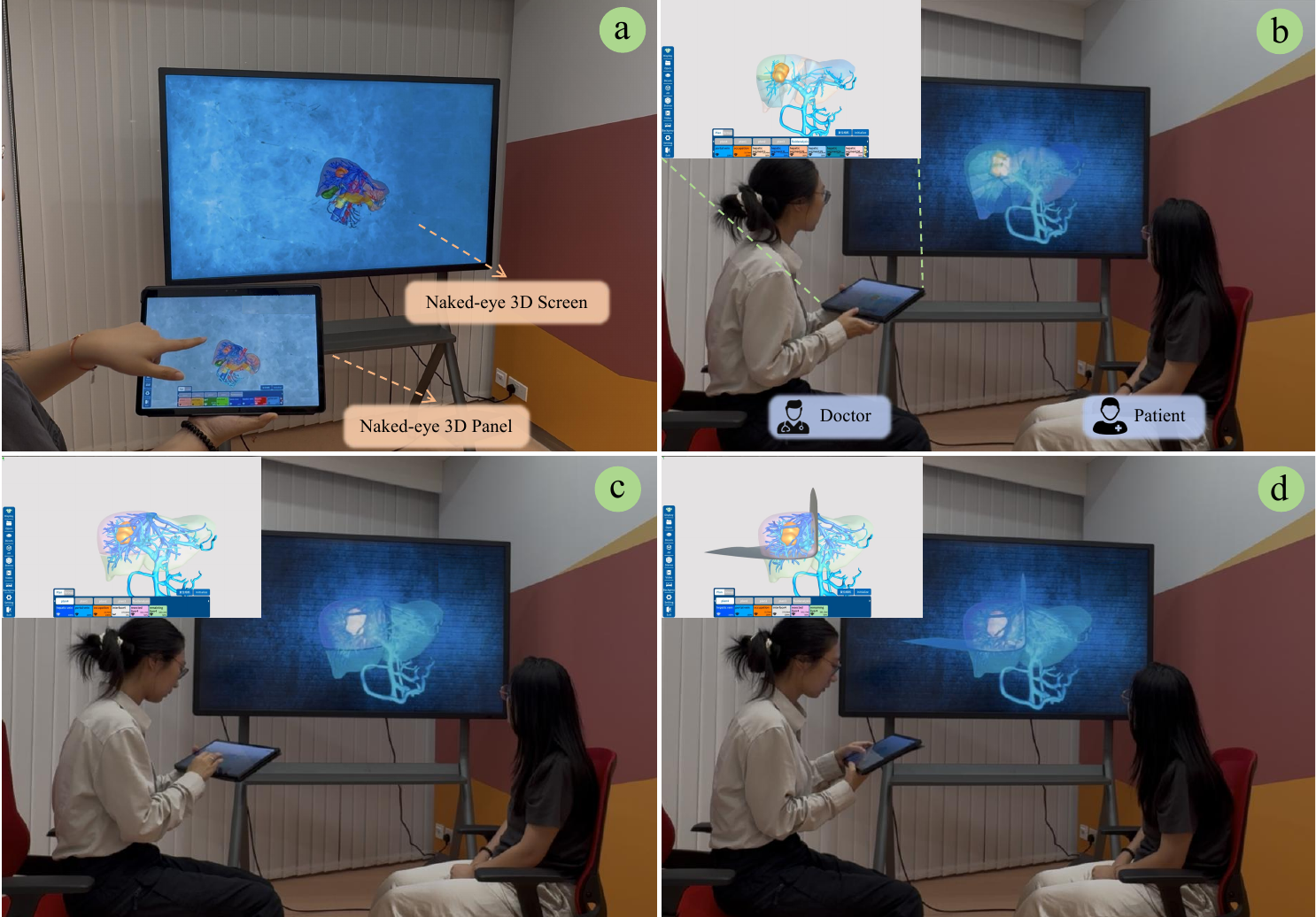} 
    \caption{Naked-eye light field 3D display platform. (a) Coordinated visualization and interaction across multiple stereoscopic devices. (b) 3D rendering of liver anatomy with eight functional segments. (c) Delineation of the planned resection volume and liver remnant. (d) Simulation of surgical resection planes.}
    \label{fig:hepatectomy_planning_single_col}
\end{figure}

\textbf{Stereoscopic Visualization.} We build a naked-eye light field 3D display platform comprising a stereoscopic display panel and a 3D standing screen, as shown in Fig.~\ref{fig:hepatectomy_planning_single_col}. The stereoscopic display panel features viewpoint tracking to provide high-resolution images and a wide field of view for individual observers. The standing screen offers both binocular and motion parallax, allowing multiple viewers to experience immersive 3D visualization simultaneously. Synchronous display and interaction between the panel and the standing screen are supported by a Unity application. This 3D display platform not only facilitates collaboration among clinicians but also enhances doctor-patient communication by presenting surgical plans in a clear and engaging way.

\section{User Experience}
We conducted user experience studies to evaluate the effectiveness of our proposed collaborative XR prototype. For surgical planning, we additionally implemented a desktop baseline using 3D Slicer~\cite{fedorov20123d}, a popular open-source software package for image analysis and scientific visualization. We invited a total of eight liver surgeons ($n=8$) to assess their experience using different systems for liver resection surgical planning. As shown in Tab.~\ref{tab:user1}, participants reported a higher System Usability Scale (SUS)~\cite{brooke1996sus} score when using our XR system (76.25$\pm$13.43) compared to the desktop counterpart (38.44$\pm$16.90), representing a +98.36\% improvement. Moreover, the participants rated their overall system experience with the XR system at an average of 4.75$\pm$0.46 on a 5-point Likert scale, indicating a high level of satisfaction.

For the stereoscopic visualization study, the participants were engineering graduate students ($n=10$) with limited medical knowledge. During the user study, a doctor demonstrator used the devices to present and explain the 3D scanning and surgical planning data of a liver cancer case. The participants were then asked to collaboratively visualize and explore the medical data using three different setups: the desktop baseline, the naked-eye 3D panel (shown in Fig.~\ref{fig:hepatectomy_planning_single_col}), and the coordinated platform (\emph{i.e.}, the naked-eye 3D panel combined with a 3D screen). 
Results are shown in Tab.~\ref{tab:user2}. A one‑way repeated‑measures ANOVA revealed significant differences in SUS ($F(2,18)=4.119, p=0.034$). Post-hoc paired t-tests further suggested that the coordinated platform's SUS was significantly higher than that of the desktop platform ($p=0.036$). The results indicate a more effective and intuitive medical collaboration experience on the coordinated platform.

\begin{table}
  \caption{User experience study of immersive XR surgical planning. The study ($n=8$) comprises 7 male and 1 female participants with an average age of 27.88. All participants are liver surgeons with limited XR experience.}
  \label{tab:user1}
  \centering%
  \renewcommand{\arraystretch}{1.1} 
  \resizebox{.85\linewidth}{!}{
  \begin{tabu}{%
    c|cc|c
    }
  \toprule
  Metric &Desktop Baseline &Immersive XR System & Improvement\\\midrule
    \textbf{System Usability} &\multirow{2}{*}{38.44$\pm$16.90} &\multirow{2}{*}{\textbf{76.25$\pm$13.43}} &\multirow{2}{*}{\textbf{+98.36\%}}\\
    \textbf{Scale} (SUS) &&&\\
  \bottomrule
  \end{tabu}%
  }
\end{table}

\begin{table}
  \caption{User experience study of coordinated stereoscopic visualization platform. The study ($n=10$) comprises 5 male and 5 female participants with an average age of 27.50. All participants are graduate students from engineering schools.}
  \label{tab:user2}
  \centering%
  \renewcommand{\arraystretch}{1.1} 
  \resizebox{.85\linewidth}{!}{
  \begin{tabu}{%
    c|ccc
    }
  \toprule
      Metric &Desktop Baseline &3D Panel Only &Coordinated Platform\\\midrule
    \textbf{System Usability} &\multirow{2}{*}{63.60$\pm$20.70} &\multirow{2}{*}{72.20$\pm$14.50} &\multirow{2}{*}{\textbf{81.00$\pm$13.80}}\\
    \textbf{Scale} (SUS) &&&\\\midrule
    \textbf{User Satisfaction} &\multirow{2}{*}{3.40$\pm$1.08} &\multirow{2}{*}{3.80$\pm$0.92} &\multirow{2}{*}{\textbf{4.20$\pm$0.92}}\\
     (5-point Likert scale) &&&\\
  \bottomrule
  \end{tabu}%
  }
\end{table}

\section{Conclusion}
Our collaborative XR prototype shows considerable promise for both clinical and educational applications, enabling real-time surgical planning, multi-user collaboration, and interactive visualization across various XR devices. Future enhancements will focus on optimizing the user experience, expanding device compatibility, and integrating AI-assisted planning tools. In the long term, we expect that our design can benefit the medical community by integrating advanced AI and XR technologies to improve clinical collaboration and workflow efficiency.

\acknowledgments{This work was supported by the Research Grants Council of the Hong Kong Special Administrative Region, China (Project No.: T45-401/22-N).}

\bibliographystyle{abbrv-doi}

\bibliography{template}
\end{document}